\documentclass[amsmath, amssymb,10pt, aps, prl, twocolumn, notitlepage, longbibliography, superscriptaddress]{revtex4-2}

\usepackage{graphicx}
\usepackage{dcolumn}
\usepackage{bm}
\usepackage{hyperref}
\usepackage{xcolor}
\usepackage{braket}
\usepackage{tabularx}
\usepackage{booktabs}
\usepackage{multirow}
\usepackage{relsize}
\usepackage{bibunits}
\defaultbibliography{Bibliography}
\defaultbibliographystyle{apsrev4-2}

\newcommand{\bk}{{\bf k}}

\newcommand{\bR}{{\bf R}}

\newcommand{\hS}{{\hat S}}

\newcommand{\hT}{{\hat T}}

\newcommand{\dd}{\mathrm{d}}
\newcommand{\EE}{\mathrm{e}}
\newcommand{\cHd}{\hat{\mathcal{H}}_{\mathrm{disorder}}}
\newcommand{\dG}{d_{\mathrm{G}}}
\newcommand{\proj}{\hat{\mathcal{P}}}
\newcommand{\Id}{\hat{\mathbb{I}}}

\newcommand{\cD}{{\cal D}}

\newcommand{\cH}{\hat{\cal H}}

\newcommand{\cO}{{\cal O}}

\newcommand{\cF}{{\cal F}}

\begin{document}
\begin{bibunit}
\nocite{apsrev42Control}


\title{Origin of the hidden energy scale and the $f$-ratio in geometrically frustrated magnets}%

\date{\today}

\begin{abstract}
Sufficiently clean geometrically frustrated (GF) magnets
are the largest class of candidate materials that may host 
quantum spin liquids (QSLs).
Some of them have been shown to exhibit 
spin-glass freezing, potentially precluding QSLs, at the ``hidden energy scale'', which is significantly lower than the microscopic energy scale of spin interactions.
Here, we investigate the origin of the hidden energy scale and its relationship to the $f$-ratio,
the figure of merit for the degree of frustration in GF magnetic materials. 
The available experimental 
and numerical data provide evidence that GF magnets display, universally, 
two distinct temperature scales in the specific heat, the lowest of which is 
of the order of the hidden energy scale $T^*$. 
We argue that this scale is determined by 
non-magnetic excitations, similar to spin exchanges in chains of
spins.
The collective entropy of such excitations
matches the entropy of the ground states of the Ising model
on the same lattice, which provides a way to verify the proposed scenario in experiment.
We demonstrate that in the presence of quenched disorder, a broad class of materials
exhibits spin-glass freezing at temperatures of order $T^*$, in accordance with experimental observations. As $T^*$ is a property of the clean GF medium, it leads to a constraint on the $f$-ratio.
\end{abstract}

\author{Phillip Popp}
\affiliation{Physics Department, University of California, Santa Cruz, California 95064, USA}

\author{Arthur P. Ramirez}
\affiliation{Physics Department, University of California, Santa Cruz, California 95064, USA}

\author{Sergey Syzranov}
\affiliation{Physics Department, University of California, Santa Cruz, California 95064, USA}

\maketitle

Geometrically frustrated (GF) magnets are considered to be a promising platform 
for hosting coveted quantum spin-liquid (QSL) states~\cite{SavaryBalents:review}. The geometry of the
lattice combined with
antiferromagnetic interactions 
leads to competing configurations of spins in such materials,
obstructing the establishment of magnetic order and possibly
giving the magnetic state liquid-like properties.

An important figure characterizing the degree of geometric frustration is the $f$-ratio~\cite{Ramirez:fRatio}
\begin{align}
    f=\theta_W/T_c,
    \label{f}
\end{align}
where $\theta_W$ is the Weiss constant, the characteristic
energy scale of interactions between the spins in the material,
and $T_c$ is the critical temperature at which the spins
order magnetically or undergo spin-glass freezing.
On a non-frustrating or weakly frustrating lattice, the critical temperature
$T_c$ is determined solely by the interaction strength in clean materials,
and the $f$-ratio is of order unity.
By contrast, strong geometric frustration prevents magnetic order from
establishing and thus significantly lowers the ordering-transition temperature or possibly cancels the transition.
Because the $f$-ratio quantifies a material's ability to resist long-range order, 
it has been adopted as a figure-of-merit for the material's potential to become a QSL.

Quenched disorder limits the $f$-ratio and, due to possibly inducing the spin-glass state, the material's ability to become a QSL. In response to quenched disorder, 
GF magnets have recently been
shown~\cite{Syzranov:HiddenEnergy} to display several
surprising, universal trends.
Contrary to the common expectation that purifying
a material suppresses spin-glass freezing, GF magnets exhibit
the opposite trend:
decreasing the density of vacancy defects, the most common type 
of disorder in such systems, increases the critical temperature of the spin-glass (SG) transition. For vanishingly low vacancy
concentrations,
a GF material would undergo 
a spin-glass transition at a ``hidden energy scale'' $T^*\lesssim 
\theta_W/10$, 
which remains finite despite the
disappearance of the SG signal in susceptibility.
While it remains difficult to fully eliminate
quenched disorder, the purest systems 
exhibit robust signs of short-range order near $T^*$ in neutron 
scattering~\cite{Broholm:neutrons,Gardner:neutrons,StockBroholm:neutrons,Syzranov:HiddenEnergy}
and specific heat, the
quantity of interest for this work, regardless of the details of 
quenched disorder.

In this paper, we investigate the origin of the hidden energy scale and its relationship to the $f$-ratio. We demonstrate that the hidden energy scale is a property of the clean GF material and reveal its microscopic mechanism. We also argue that 
the $f$-ratio can be generalised
to include the possibility that when a material undergoes a crossover at a low-temperature
scale $T_c$, accompanied by thermodynamic deactivation of magnetic degrees of freedom, it does not exhibit a magnetic or spin-glass phase transition.




To illustrate the nature of the hidden energy scale, 
we give both experimental and theoretical
evidence that the heat capacity $C(T)$ in GF materials exhibits two 
distinguishable peaks, with the lower-temperature peak located
at temperatures of order $T^*$.

This lower-temperature 
peak comes from excitations similar to spin exchanges in chains of spins, 
whose energy 
is well separated from the energy of other excitations,
similar to spin flips,
which give rise to the higher-temperature peak in $C(T)$.
We demonstrate that in the presence of quenched disorder,
the material 
is sensitive to quenched disorder at temperatures $T\lesssim T^*$,
which is consistent with the experimentally observed SG freezing of 3D GF materials
at such temperatures. 
We also predict the entropy
of the excitations giving rise to the lower-temperature peak, which allows for further
verification of the proposed mechanism of the hidden energy scale.



{\it Evidence for two distinct temperature scales in frustrated magnets.}
Key to the proposed scenario of the hidden energy scale 
is our observation, discussed in this paper, 
that GF magnets that lack long-range magnetic order at low temperatures universally exhibit two 
distinct temperature scales in the dependence
of $C(T)$  on temperature.

\begin{figure}
    \centering
    \includegraphics[width=0.45\textwidth]{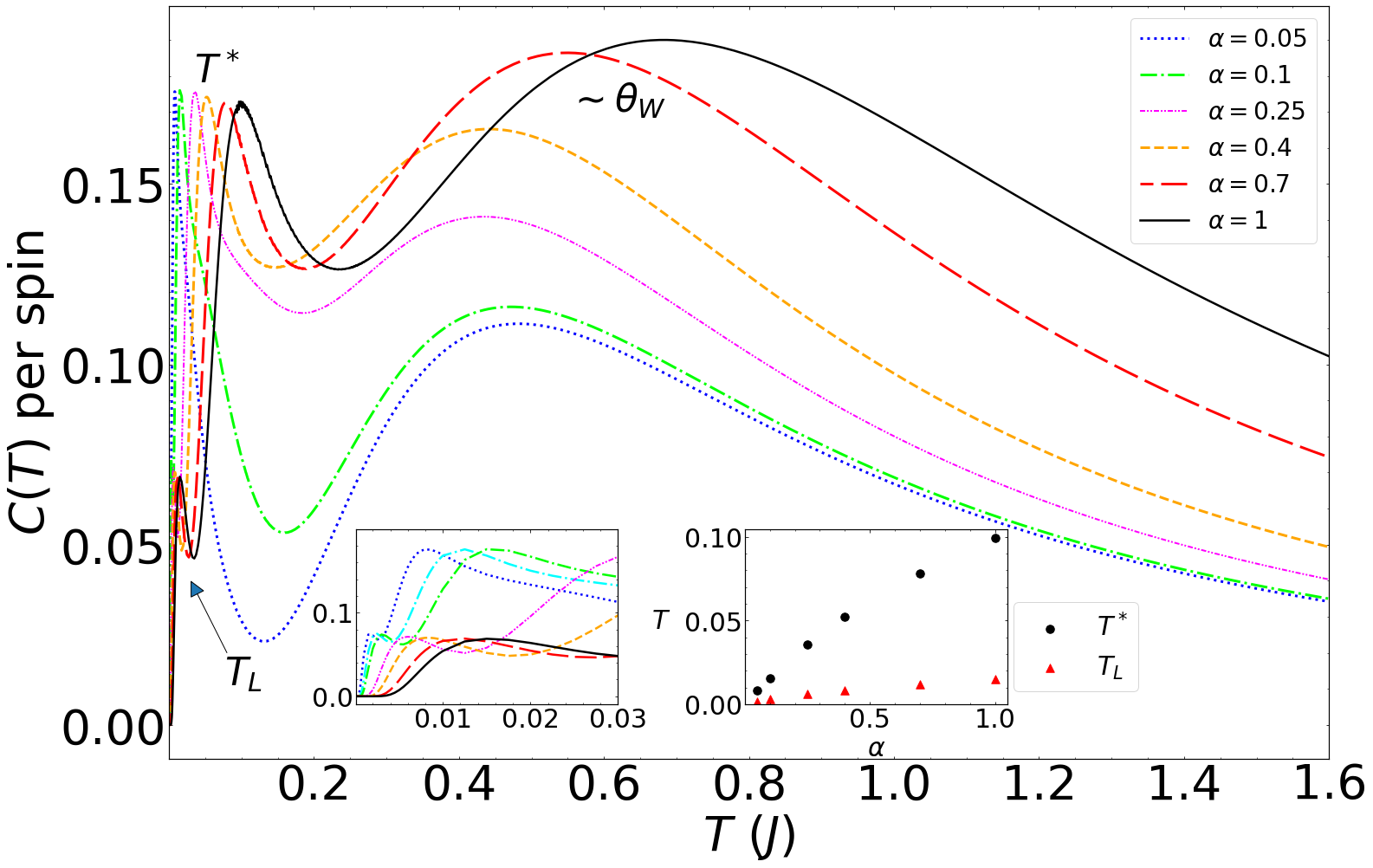}
    \caption{
    Heat capacity $C(T)$ of the spin-1/2 $XXZ$ model on a cluster of 18 sites on the kagome lattice
    computed from exact diagonalization with periodic boundary conditions
    for various values of the anisotropy parameter $\alpha$ [cf. Eq.~\eqref{eq:XXZ}].
    The left inset shows $C(T)$
    at very low temperatures, which
    may display a secondary low-temperature peak. The right inset shows the $\alpha$-dependence of the primary {($T^*$)} and secondary {($T_L $)} low-temperature peak positions.
    The existence of distinguishable low- and high-temperature
    peaks in $C(T)$ is in agreement with the results of Refs.~\cite{Elser:KHAF,ElstnerYoung:kagome,IsodaNakano:XXZ,SugiuraShimizu:KHAF,Munehisa:KHAF,ShimokawaKawamura:KHAF,SindzingreMisguich:KHAF,ZengElser:KHAF,MisguichBernu:KHAF,NakamuraMiyashita:KHAF,TomczakRichter:KHAF,MoritaTohyama:KagomeTriangular,Ulaga:EasyAxis} for the isotropic Heisenberg model ($\alpha=1$)
    and of Refs.~\cite{IsodaNakano:XXZ,Ulaga:EasyAxis} for the $XXZ$ models. 
    As we clarify in the Suppmental Material~\cite{SM},
    the peaks may merge and turn into a plateau for $\alpha\sim 1$
    and other boundary conditions.
    }
    \label{fig:KHAF}
\end{figure}

In experiments~\cite{Greywall:DoublePeakHe,Schiffer:GGG,Ishida:HeRingExchange,NakatsujiNambu:THAF,Li:S2TAF,Bordelon:NaYbO2,Ranjith:NaYbSe2,CuGa2O4:ToUpdate} on GF materials, in which the data extend 
over a large enough temperature range,
these two temperature scales manifest themselves 
in two distinct, well-distinguishable peaks in $C(T)$.
This behaviour is mirrored by numerical simulations~\cite{Elser:KHAF,ZengElser:KHAF,ElstnerYoung:kagome,NakamuraMiyashita:KHAF,TomczakRichter:KHAF,SindzingreMisguich:KHAF,MisguichBernu:KHAF,MisguichSindzingre:KHAF,IsodaNakano:XXZ,SugiuraShimizu:KHAF,Munehisa:KHAF,ShimokawaKawamura:KHAF,SchnackSchulenberg:KHAF,PrelovsekKokalj:THAF,ChenQu:THAF,MoritaTohyama:KagomeTriangular,SekiYunoki:RingExchange,HutakKrokhmalskii:HKHAF,Ulaga:EasyAxis} on small clusters
of spins on frustrating lattices, exemplified 
by spins on the kagome lattice, whose heat capacity is shown
in Fig.~\ref{fig:KHAF} (where the two peaks are labelled as
$T^*$ and $\sim\theta_W$).

For some of the numerically simulated systems, the lower-temperature
peak is less distinguishable from the larger, higher-temperature peak and forms a ``shoulder'' in its vicinity~\cite{SugiuraShimizu:KHAF,Munehisa:KHAF,ChenQu:THAF,PrelovsekKokalj:THAF,SchnackSchulenberg:KHAF,Gonzalez:THAF}.
We note that the resolution of numerical simulations may be limited by finite-size effects. Given, however, that the peaks are readily distinguishable in all experiments
and given the preponderance of numerical evidence, the existence of two temperature scales in $C(T)$
is a universal feature of GF materials.
In Table~\ref{tab:C}, we summarise the available
experimental and numerical 
evidence~\cite{pyrochlorenote}.

\begin{table}[t]
    \centering
    \begin{ruledtabular}
    \begin{tabular}{c|c}
         Lattice & System and references\hspace*{1.4cm}\\
         \hline \\
         \multirow{2}{*}{Kagome}& $^3He$ on graphite~\cite{Greywall:DoublePeakHe,Ishida:HeRingExchange}\hspace*{1.4cm} \\
           & Numerics~\cite{Elser:KHAF,ElstnerYoung:kagome,IsodaNakano:XXZ,SugiuraShimizu:KHAF,Munehisa:KHAF,ShimokawaKawamura:KHAF,SchnackSchulenberg:KHAF,MisguichSindzingre:KHAF,SindzingreMisguich:KHAF,ZengElser:KHAF,MisguichBernu:KHAF,NakamuraMiyashita:KHAF,TomczakRichter:KHAF,MoritaTohyama:KagomeTriangular,Ulaga:EasyAxis}, this work\hspace*{1.4cm} \\
         \hline \\
        \multirow{3}{*}{Triangular}&$NiGa_2S_4$~\cite{NakatsujiNambu:THAF}, $FeAl_2Se_4$~\cite{Li:S2TAF},\hspace*{1.4cm} \\
        &$NaYbO_2$~\cite{Bordelon:NaYbO2}, $NaYbSe_2$~\cite{Ranjith:NaYbSe2},\hspace*{1.4cm} \\&Numerics~\cite{ChenQu:THAF,PrelovsekKokalj:THAF,SekiYunoki:RingExchange,Ulaga:EasyAxis}\hspace*{1.4cm}
         \\
         \hline \\
         Hyperkagome &
         $Gd_3 Ga_5 O_{12}$~\cite{Schiffer:GGG}\hspace*{1.4cm}\\ &Numerics~\cite{HutakKrokhmalskii:HKHAF}\hspace*{1.4cm} \\
         \hline \\
         Spinel       & $CuGa_2O_4$~\cite{CuGa2O4:ToUpdate}\hspace*{1.4cm}  \\ 
    \end{tabular}
    \end{ruledtabular}
    \caption{
    Experimental and numerical evidence for two temperature scales in systems
    of spins on various geometrically frustrating lattices.
    }
    \label{tab:C}
\end{table}

\begin{figure*}
    \centering
    \includegraphics[width=0.9\textwidth]{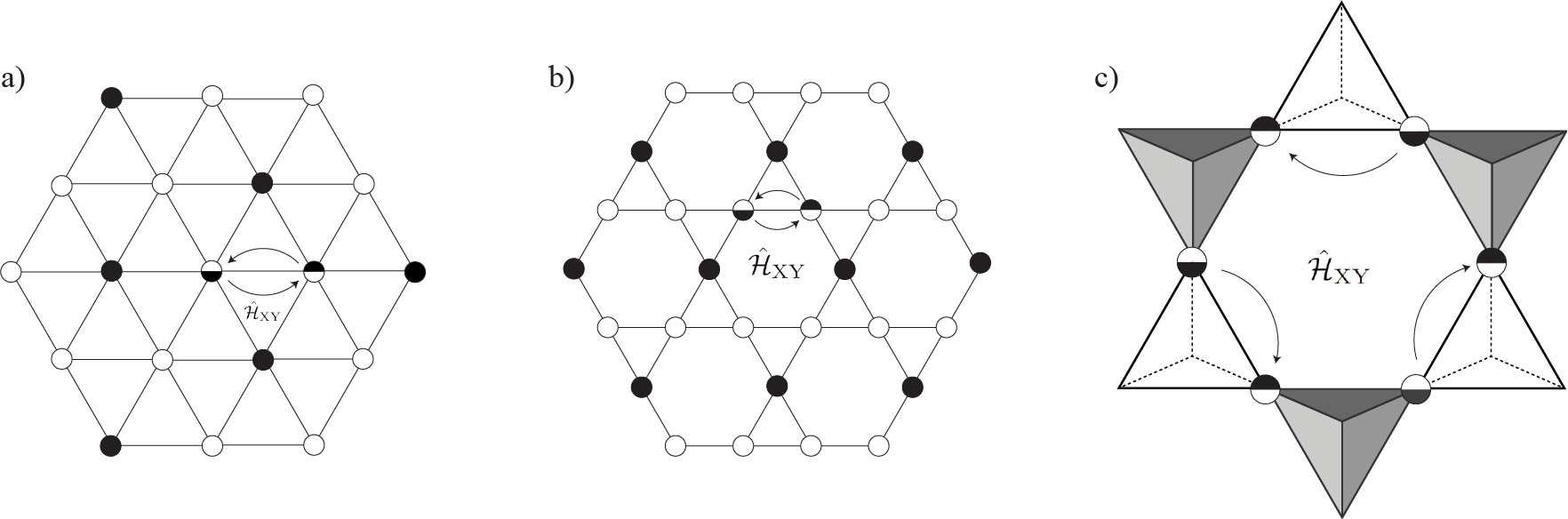}
    \caption{Illustrations of Ising ground state hyridization via the transverse coupling $\cH_{\text{XY}}$ on the a) triangular, b) kagome, and c) pyrochlore lattices. Each picture shows
    two Ising ground states which are hybridized by $\cH_{\text{XY}}$. Fully-filled (open) circles indicate spin-up (down) on the corresponding sites, and are the same between the two states. Half-filled circles 
    correspond to the upper spin orientations in one Ising state and the lower orientations in the other.
    }
    \label{fig:hybridization}
\end{figure*}

{\it Excitations near the hidden energy scale.}
The higher energy scale in $C(T)$ is of
order of the Weiss temperature 
\begin{align}
\theta_W\sim ZJ
\label{WeissT}
\end{align}
in each material,
where $Z$ and $J$ are, respectively, the coordination number
of the magnetic sublattice of the material
and the characteristic exchange coupling. The scale $\theta_W$ can be qualitatively 
understood as the energy of flipping a spin interacting with $Z$ neighbours
with the interaction energy $J$. Although we focus, for simplicity,
on spins-$1/2$ hereafter,
our results hold for arbitrary spins and are consistent with the range of experimental examples.

To illustrate the origin of the lower energy scale, we consider the Hamiltonian 
\begin{subequations}
\begin{align}
    &\cH=
    \cH_\text{Ising}
    +\cH_\text{XY}
    +\cH_\text{disorder}+\cH_\text{weak}
        \label{eq:XXZ}
    \\
    &\cH_\text{Ising}=
    J\sum_{(ij)}\hS_i^z\hS_j^z
    \label{eq:XXZ-XY}
    \\ 
    &\cH_\text{XY}=
    \alpha J\sum_{(ij)}\left(\hS_i^x\hS_j^x+\hS_i^y\hS_j^y\right)
        \label{eq:XXZc}
    \\
    &\cHd = \sum_{(ij)} \delta J_{ij} \hS_i^z \hS_j^z
    \label{eq:zzfluctuations}
\end{align}
\end{subequations}
of spins on a GF lattice, where
the part $\cH_\text{Ising}
    +\cH_\text{XY}$ describes a Heisenberg $XXZ$ model;
    the summation is carried out over all pairs $(ij)$ of neighbouring spins;
    $J$ is the coupling strength between the spins' $z$ components;
$\cH_\text{disorder}$ is a perturbation created by weak quenched disorder whose exact form is not important but is considered, for concreteness, to come from the random fluctuations $\delta J_{ij}$ of the exchange couplings; $\cH_\text{weak}$ accounts for weak non-exchange, e.g. dipole-dipole, interactions with characteristic energies significantly
smaller than $J$, and the anisotropy parameter $\alpha$ describes the relative strength of
the transverse and longitudinal exchange interactions.

At $\alpha=0$ (the Ising limit) and $\cH_\text{disorder}
=\cH_\text{weak}=0$, spins on a GF lattice have
extensive degeneracy, i.e. degeneracy that 
scales exponentially with the system size, whereas single spin-flip excitations have a
characteristic energy of the order of $\theta_W\sim JZ$.

For $0< \alpha\ll 1$,
the degeneracy of the ground states is lifted by disorder and the transverse interactions,
but the energies of respective low-lying excitations remain well separated from the characteristic energies of the excitations of the Ising model.
Thus, at small $\alpha$ and weak disorder and other interactions, the model exhibits two types of excitations,
adiabatically connected to, respectively, the ground states and the excited states of the Ising model,
which gives rise to the two peaks in the heat capacity $C(T)$.
We argue below that for most materials, the two peaks remain well separated for $\alpha\sim 1$,
including in the isotropic Heisenberg models.

Indeed, the hybridisation of the ground states of the Ising model
at non-zero $\alpha$ is caused by the transverse-coupling Hamiltonian $\cH_\text{XY}$ given by Eq.~\eqref{eq:XXZc}.
The $n$-th order transition amplitude between two Ising ground states $\ket{G_1}$
and $\ket{G_2}$ can be estimated as
\begin{align}
    T_{G_1G_2}\sim 
    \bra{G_1}
    \cH_\text{XY}\left(\hat{\cal \tilde{H}}_\text{Ising}^{-1}\cH_\text{XY}\right)^{n-1}
    \ket{G_2},
    \label{TransitionAmplitude}
\end{align}
where $\hat{\cal \tilde{H}}_\text{Ising}$ is the Ising Hamiltonian 
$\hat{\cal {H}}_\text{Ising}$ in the reduced space of the excited states of the Ising model.
The minimum order $n$ of the processes 
hybridizing the Ising ground states
is given by the minimum number $Z_1$ of pairwise exchange processes of nearest-neighbour
antiparallel spins that transform one Ising ground state to another.

Examples of such ``Ising instanton'' processes are shown in Fig.~\ref{fig:hybridization}.
On the triangular and kagome lattices, $Z_1=1$,
corresponding to the minimum-exchange processes between possible
ground states showing in
Figs.~\ref{fig:hybridization}a and \ref{fig:hybridization}b.
For the pyrochlore lattice, $Z_1=3$, corresponding to 
spin exchange in a loop of $6$ spins~\cite{HermeleFisherBalents:photons},
as shown in Fig.~\ref{fig:hybridization}c.

The characteristic eigenvalues of the operator $\hat{\tilde{H}}_\text{Ising}$ in Eq.~\eqref{TransitionAmplitude} are given by $\theta_W\sim ZJ$,
while the matrix elements of the operators 
$\cH_\text{XY}$ are given by $\alpha J/2$.
The transition amplitude between the Ising ground states due to
$Z_1$ pairwise spin-exchange processes
can, therefore, be estimated as 
\begin{align}
    T^*\sim \left[\alpha J/\theta_W\right]^{Z_1-1}\alpha J
    \sim J \alpha^{Z_1}/Z^{Z_1-1}.
    \label{BeyonceScale}
\end{align}
The quantity~\eqref{BeyonceScale} describes the characteristic 
energy splitting between the lowest energy levels of the $XXZ$
model 
with the Hamiltonian~\eqref{eq:XXZ}-\eqref{eq:XXZc} in the limit of weak quenched disorder and 
weak non-exchange interactions.
These states are adiabatically connected to the Ising 
ground states on the same lattice when the parameter 
$\alpha$ increases from $0$ to a nonzero value.
The quantity $Z_1$ in Eq.~\eqref{BeyonceScale} is given by the minimum 
number of pairwise spin-exchange processes that can connect two Ising ground states. The value of $Z_1$, like $Z$, depends on the details of the lattice.

{\it Excitations on the kagome lattice.} 
Numerical simulations~\cite{IsodaNakano:XXZ,Ulaga:EasyAxis}
of the $XXZ$ Heisenberg models 
on the kagome lattice confirm the persistence of the two peaks up to $\alpha\sim 1$.
To illustrate this trend, we show in Fig.~\ref{fig:KHAF} the heat 
heat capacity of a cluster of $18$ spins
on the kagome lattice for various values of $\alpha$ that we obtain
by exact diagonalization. 
The temperatures $T^*$ and $T_L$ of the two low-temperature peaks,
shown in the inset in Fig.~\ref{fig:KHAF}, scale $\propto \alpha$ at small $\alpha$,
which demonstrates that the associated excitations at $\alpha\sim 1$
are adiabatically connected to single spin-exchange processes {($Z_1=1$)}
at $\alpha=0$ shown in Fig.~\ref{fig:hybridization}b.
We note the emergence of the small ``secondary'' peak at
temperature $T_L$.
Such a feature was also identified for the isotropic Heisenberg model ($\alpha=1$) in Refs.~\cite{SindzingreMisguich:KHAF,Munehisa:KHAF,ShimokawaKawamura:KHAF,SchnackSchulenberg:KHAF,MoritaTohyama:KagomeTriangular}, where it was suggested
to be a finite-size effect.

{\it Constraint on the $f$-ratio.}
The scales~\eqref{WeissT} and \eqref{BeyonceScale} in 
a generic GF material determine the characteristic
energies of the excitations that give rise to the two peaks in the  behaviour of the heat
capacity $C(T)$.
As GF lattices typically have coordination numbers in the
interval $Z=4\ldots6$ and the quantity $Z_1=1\ldots3$, the ratio
\begin{align}
    f=\frac{\theta_W}{T^*}\sim \left(\frac{Z}{\alpha}\right)^{Z_1}
    \label{fcon}
\end{align}
of the two scales
is large even when the parameter $\alpha$ is not small,
resulting in a separation of the peaks of $C(T)$ in realistic materials. 
{The order of magnitude of the ratio of the 
temperatures of higher and lower peaks, whose detailed analyses will
be presented elsewhere, is consistent with the estimate~\eqref{fcon}.}

At temperatures $T\lesssim T^*$, the excitations that give rise to the lower-temperature
peak are suppressed, and the magnetic degrees of freedom of the material freeze.
The hidden energy scale $T^*$, thus, constrains the 
$f$-ratio in GF magnets.

{\it Spin-glass freezing in the presence of disorder.} In 3D systems with quenched disorder,
the thermodynamic freezing of the magnetic degrees of freedom at
$T\lesssim T^*$ will lead to a spin-glass-freezing transition
at temperatures $T_c\sim T^*$ for a broad range of disorder strengths.

To demonstrate this, we compute (see Supplemental Material~\cite{SM}
for details) the critical value of the variance $\varkappa_c=
\left<\delta J_{ij}^2\right>$ of the fluctuations $\delta J_{ij}$ 
of the exchange couplings in the mean-field replica-symmetric approximation~\cite{SherringtonKirkpatrick:SG,Binder:SG,Nishimori:SG} assuming
that the fluctuations on different bonds are independent.
Up to a non-universal prefactor that depends on the details of the lattice, the 
critical disorder strength of the glass transition is given by
\begin{equation}
    \varkappa_c (T) \sim
    \begin{cases}
    &\left[ Z \mathlarger{\int}_{\substack{\text{lower} \\ \text{peak}}} \frac{\rho (\xi)}{\xi^2} \, \dd \xi \right]^{-1}
    = A (T^*)^2/Z, \quad T \ll T^* \\
    &T^2/Z, \quad T^* \ll T \ll \theta_W,
    \end{cases}
    \label{eq:J0limits}
\end{equation}
where the integral is taken over the low-energy states, i.e. states 
that give rise to the low-temperature peak of $C(T)$; $\rho(\xi)$
is the density of such states in the clean system, and 
$A$ is a dimensionless constant.

As we discuss in the Supplemental Material~\cite{SM}, the constant $A$ in Eq.~\eqref{eq:J0limits}
may be either zero or of order unity, depending on the excitation energy spectrum in a particular system, whose detailed investigation for specific materials we leave for future studies. In both cases, however, the critical disorder strength $\varkappa_c$ rapidly grows at temperatures $T\gtrsim T^*$.
This suggests that the spin-glass transition will be observed at temperatures
$T\sim T^*$ for a broad range 
of disorder strengths, in agreement with the recent analyses of spin-glass transitions
in GF magnets~\cite{Syzranov:HiddenEnergy}.

{\it The two-peak structure of heat capacity is specific to GF materials.}
We emphasise that non-GF materials will not
display the two
distinct energy scales we have described
unless the spin-spin interactions contain two distinct energies~\cite{deJonghMiedema:review}.
Indeed, the ground states of a model that develops long-range magnetic order at low
temperatures will not have extensive degeneracy. The excitations that are adiabatically connected to the Ising ground states
when $\alpha$ is changed from $\alpha=0$ to a nonzero value will, therefore,
have a vanishing contribution to $C(T)$ per spin and will not lead to the formation of 
the low-temperature peak.

\begin{table}[hb]
    \centering
    \begin{ruledtabular}
    \begin{tabular}{c|c}
         Lattice & Approximate entropy per site\hspace*{1.8cm} \\
         \hline \\
         Kagome & $0.50183$ (\cite{KanoNaya:KIAF,SinghRigol:KIAF})\hspace*{1.8cm} \\
         \hline \\
         Triangular & $0.323066$ (\cite{Wannier:TIAF,HwangKim:TIAF,Kim:TIAF})\hspace*{1.8cm} \\
         \hline \\
         Pyrochlore & $0.203$ (\cite{RamirezHayashi:SpinIce,Siddharthan:Pyrochlore,LauFreitas:SpinIce,SinghOitmaa:PIAF,Lin:PIAF})\hspace*{1.8cm} 
    \end{tabular}
    \end{ruledtabular}
    \caption{
    Entropies of the ground states of the Ising model on various lattices.
    }
    \label{tab:entropy}
\end{table}

{\it Experimental test: entropy related to the ``hidden energy scale''.} In materials with well-separated peaks of $C(T)$ (see, e.g. Refs.~\cite{Greywall:DoublePeakHe,NakatsujiNambu:THAF,CuGa2O4:ToUpdate}), the entropy $S_\text{low}=\int_{\substack{\text{lower} \\ \text{peak}}} \frac{C(T)}{T} \mathrm{d} T$ associated with the lower peak is a well-defined 
quantity intrinsic to the pure material. Because the excitations that give rise to 
that peak are adiabatically connected to the ground states of the Ising model on the same lattice, the entropy 
$S_\text{low}$ matches the entropy of the ground 
states of that Ising model. The value of this entropy also remains unaltered by weak interactions and disorder so long as the peaks of $C(T)$ are well separated.

For common $2$D lattices, the values of the respective Ising ground-state
entropies can be obtained analytically, while for 
Ising models on other lattices, the entropy values are known from numerical 
simulations, as summarised in Table~\ref{tab:entropy}. Because the values 
of those (Ising) entropies match the entropies corresponding to the lower peaks 
in the more complicated, realistic model with the Hamiltonian~\eqref{eq:XXZ}-\eqref{eq:zzfluctuations},
the value of that entropy of either peak of $C(T)$ can
be used as an experimental test of the origin 
of the hidden energy scale.

This is exemplified by the layered triangular-lattice
compound $NiGa_2S_4$, for which the lower peak entropy obtained from the digitized data of Ref.~\cite{NakatsujiNambu:THAF} is $S \approx 0.35$ per spin and is close to the Ising ground-state entropy $S_\triangle\approx 0.32$ for the triangular lattice~\cite{Wannier:TIAF,HwangKim:TIAF,Kim:TIAF}.
If the low-temperature 
behaviour of $C(T)$ also exhibits a secondary low-temperature peak (cf. Fig.~\ref{fig:KHAF}), the entropy under this secondary peak should
be included in the respective entropy of the lower-temperature peak.
{The ratio $f^\prime \approx 8$ of the temperatures of the higher- and lower-temperature peaks in $NiGa_2S_4$ is close to the estimate
$f=6$ given by Eq.~\eqref{fcon}}.

{\it Which peak is usually visible in experiments?}
Often, due to insufficient temperature range,
only one of the two peaks is seen in experiments 
and it is not always clear which peak this is, the low- or the high-temperature one.
Our thinking in this paper
is focused on data for GF magnets containing $3d$ elements with $\theta_W$ values comparable to or greater than $100 K$.
As a result, there are few $C(T)$ measurements that capture this temperature region and the high-temperature peak
 (an exception is $NiGa_2S_4$,
with $\theta_W = 80 K$~\cite{NakatsujiNambu:THAF}). 

By contrast, in materials with low $\theta_W$, such as 
rare-earth frustrated materials, the higher-temperature peak is usually observed. For example, in $YbMgGaO_4$ ($\theta_W$ = 4 K)~\cite{Paddison:YbMgGaO} and $Ce_2Zr_2O_7$ ($\theta_W$ =0.6 K)~\cite{Gao:Ce2Zr2O7}, the peaks in $C(T)$
are observed, respectively,
at $2K$ and $0.2K$. 
For such low temperatures of the higher-temperature peaks,
the lower-temperature peaks should be expected at temperatures $T^*$ below 
the practical lower limit of most ultra-low-temperature experiments
(around $50 mK$).


{\it Effect of the magnetic field.} 
The low-temperature peak in the behaviour of $C(T)$ will be sensitive
or insensitive to the external magnetic field, depending on the lattice
and the range of interactions.
For some lattices, such as the pyrochlore lattice~\cite{IsakovMoessnerSondhi:pyrochlorew=},
the ``bilayer kagome'' lattice of $SrCr_8Ga_4O_{19}$~\cite{Ramirez:SCGOentropy} and the 
hyperkagome lattice of $Na_4Ir_3O_8$~\cite{OkamotoTakagi:NaIrOheat,HopkinsonKim:NaIrOstates},
the Ising ground states have zero magnetization if the exchange interactions between spins are nearest-neighbour.
As a result, all the low-energy states of the 
Heisenberg models that are adiabatically connected to those Ising states
have zero magnetizations. For this reason, the low-temperature
peak of $C(T)$ in these materials is independent of the magnetic
field, as observed in experiments~\cite{Ramirez:SCGOentropy,OkamotoTakagi:NaIrOheat}.

By contrast, if the Ising ground states allow for finite magnetizations,
exemplified by the triangular and kagome lattices with nearest-neighbour
interactions, the low-energy excitations of the corresponding Heisenberg
models will, in general, have spins. Such excitations
have been reported for the kagome lattice in Refs.~\cite{WaldtmannEverts:KHAF,SindzingreMisguich:KHAF, IsodaNakano:XXZ,Lauchli:KHAF,SchnackSchulenberg:KHAF,Ulaga:EasyAxis}.
The low-temperature peaks in such materials
will be sensitive to the magnetic field.


{\it Conclusion.} In conclusion, we argue, based on the analysis
of available experimental and numerical data, that the temperature
dependence of the heat capacity $C(T)$ exhibits two distinct temperature
scales. The order of magnitude of the lower scale matches the  
``hidden energy scale'' below which the cleanest of GF magnetic materials
show spin-glass freezing and come from excitatinos
adiabatically connected to the Ising ground states on the same lattice.
The entropy associated with the lower-temperature peak matches
the Ising ground-state entropy, which can be
used to verify our predictions experimentally.
As the value of the hidden energy scale is a property of the 
clean GF medium, it constrains the $f$-ratio, characterizing 
the degree of frustration.

{\it Acknowledgments.} We thank V. Elser for useful feedback on the manuscript and P.~Prelov\u{s}ek and M. Ulaga for useful discussions. We thank an anonymous referee for bringing Ref.~\cite{Li:S2TAF} to our attention.
Our work has been supported by the NSF grant DMR2218130
and the Committee on Research at the University of California Santa Cruz.


\putbib
\end{bibunit}


\newpage
\begin{bibunit}
\nocite{apsrev42Control}
\onecolumngrid
\vspace{2cm}

\cleardoublepage

\setcounter{page}{1}

\renewcommand{\theequation}{S\arabic{equation}}
\renewcommand{\thefigure}{S\arabic{figure}}
\renewcommand{\thetable}{S\arabic{table}}
\renewcommand{\thetable}{S\arabic{table}}
\renewcommand{\bibnumfmt}[1]{[S#1]}
\renewcommand{\citenumfont}[1]{S#1}

\setcounter{equation}{0}
\setcounter{figure}{0}
\setcounter{enumiv}{0}

\begin{center}
	\textbf{\large Supplemental Material for \\
		``Origin of the hidden energy scale and the $f$-ratio in geometrically frustrated magnets''
	}
\end{center}

\section{Boundary conditions and the heat capacity of spins on the kagome lattice}

\begin{figure*}[h]
    \centering
    \includegraphics[width=0.6\textwidth]{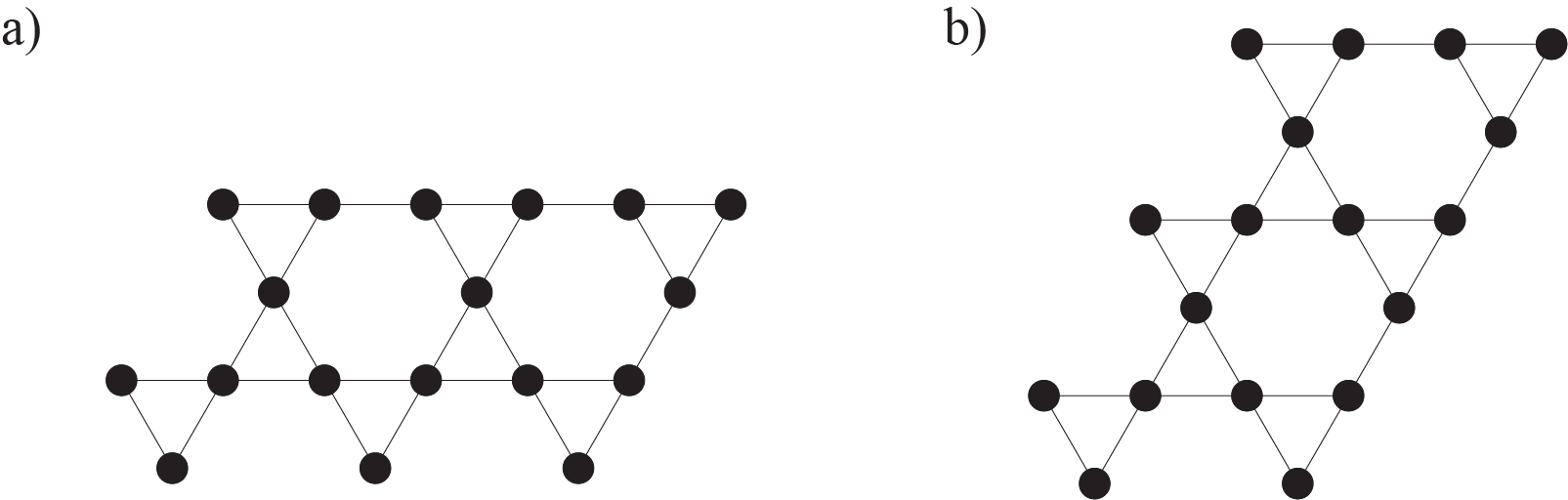}
    \caption{Two inequivalent $18$-site clusters on the kagome lattice. With periodic boundary conditions, cluster (a) has a threefold translation symmetry along the horizontal, while cluster (b) has a threefold translation symmetry along $60^\circ$ from the horizontal.}
    \label{fig:Kagome_Clusters}
\end{figure*}

\begin{figure*}[h]
    \centering
    \includegraphics[width=0.6\textwidth]{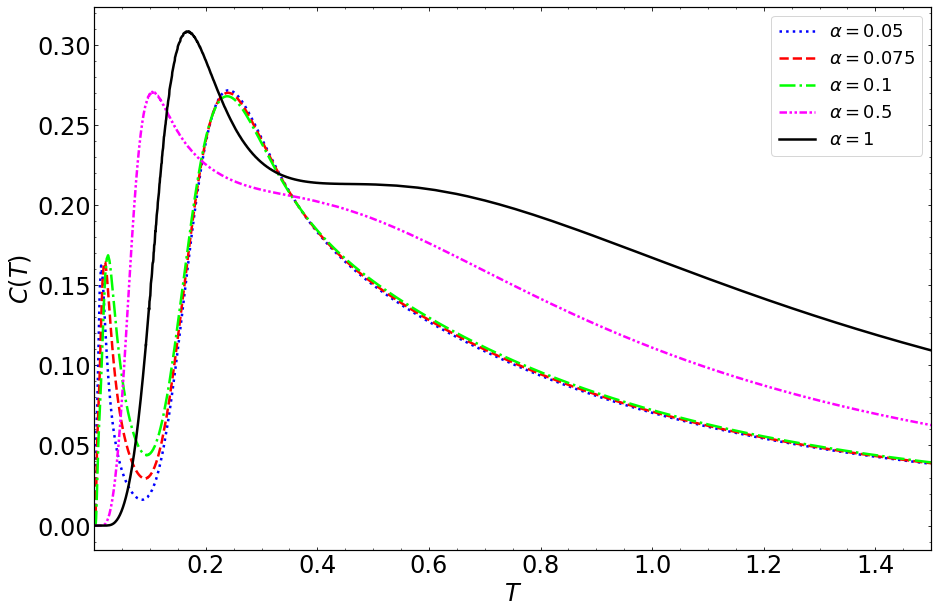}
    \caption{Heat capacity $C (T)$ (per spin) of the spin-$1 / 2$ $XXZ$ model on cluster (b) of Fig. \ref{fig:Kagome_Clusters}. For large values of $\alpha$, there are not two well-separated peaks, in contrast to heat capacities on cluster (a) and in Refs. \cite{IsodaNakano:XXZ,Ulaga:EasyAxis}.}
    \label{fig:SI_Heat_Capacities}
\end{figure*}

In this section, we investigate the dependence of the numerical results 
for $C(T)$ on the boundary conditions of spin clusters.
For the small clusters considered in Refs.~\cite{Elser:KHAF,ElstnerYoung:kagome,IsodaNakano:XXZ,SugiuraShimizu:KHAF,Munehisa:KHAF,ShimokawaKawamura:KHAF,SchnackSchulenberg:KHAF,MisguichSindzingre:KHAF,SindzingreMisguich:KHAF,ZengElser:KHAF,MisguichBernu:KHAF,NakamuraMiyashita:KHAF,TomczakRichter:KHAF,MoritaTohyama:KagomeTriangular,Ulaga:EasyAxis} and in this work,
the form of $C (T)$ is sensitive to the boundary conditions. This is exemplified by the two inequivalent $18$-site clusters of the kagome lattice shown in Fig.~\ref{fig:Kagome_Clusters}. The results we report in Fig.~\ref{fig:KHAF} were obtained using cluster (a), for which we obtain well-separated peaks in the heat capacity $C (T)$ in agreement with Refs.~\cite{IsodaNakano:XXZ,Ulaga:EasyAxis}. However, while $XXZ$ Hamiltonians with small values of $\alpha$ on cluster (b) also display two well-separated peaks in $C (T)$, for larger values of $\alpha$, up through the isotropic Heisenberg case ($\alpha=1$), the high-temperature peak merges with the low-temperature
one, forming a plateau rather than a distinct peak.
These behaviours are shown in Fig.~\ref{fig:SI_Heat_Capacities}.

While small spin clusters on the kagome lattice
may not display good separation of the peaks in the behaviour of $C(T)$,
the two peaks are rather distinct in the available
experiments~\cite{Greywall:DoublePeakHe,Schiffer:GGG,Ishida:HeRingExchange,NakatsujiNambu:THAF,Bordelon:NaYbO2,Ranjith:NaYbSe2,CuGa2O4:ToUpdate} on GF materials,
which suggests that the separation improves in the limit of 
large system sizes, which are currently inaccessible by numerical methods.
A more thorough understanding of the effects of boundary conditions and system size on the spectra and thermodynamic properties of kagome clusters awaits future study.

\section{Derivation of the critical disorder strength for glass freezing}
In this section, we present 
a derivation of the phase boundary of the spin-glass state 
in the model of a GF magnet considered in the main text.
For simplicity, we consider nearest-neighbour bond disorder in this model.
The Hamiltonian of the system is given by
\begin{subequations}
\begin{align}
    &\cH = \cH_0 + \cHd
    \label{eq:latHam}
    \\
    &\cHd = \sum_{(ij)} \delta J_{ij} \hS_i^{z \gamma} \hS_j^{z \gamma}
    \label{eq:latHamdis}
\end{align}
\end{subequations}
where $\cH_0$ is the Hamiltonian of $R$ clean, independent replicas of the system; $\cHd$ describes quenched fluctuations of the $zz$-couplings and $\gamma$ is the replica index. In the main text, the disorder-free Hamiltonian of the system is given by $\cH_{\mathrm{Ising}} + \cH_{\mathrm{XY}}$ [see Eqs.~\eqref{eq:XXZ-XY} and~\eqref{eq:XXZc}] in a single replica; however, the exact form of $\cH_0$ is not important for the present calculation. We consider each replica to consist of a finite lattice of $N$ sites with periodic boundary conditions and coordination number $Z$. The sum in Eq.~\eqref{eq:latHamdis} runs over all nearest-neighbor site pairs $(ij)$. Hereinafter, summation over repeated indices is implied.


Using the interaction representation, where we take Eq.~\eqref{eq:latHamdis} as a perturbation to $\cH_0$, we can write the partition function of the $R$ replicas for a particular realization of the bond fluctuations in the form
\begin{equation}
Z^R [\{ \delta J_{ij} \}] = Z_0^R \biggl{\langle} \hT_\tau \, \mathrm{exp} \biggl{[} -\int_0^\beta \dd \tau \, \sum_{(ij)} \delta J_{ij} \hS_i^{z \gamma} (\tau) \hS_j^{z \gamma} (\tau) \biggr{]} \biggr{\rangle}
\label{eq:ZR}
\end{equation}
where $\beta$ is the inverse temperature, $Z_0$ is the partition function of a single clean replica, $\hS_i^{z \gamma} (\tau)$ is the spin operator in the interaction representation, $\hT_\tau$ is the Matsubara-time-ordering operator, and the angle brackets denote equilibrium thermal averaging with respect to $\cH_0$.

Assuming, for simplicity, that the fluctuations are Gaussian-distributed independently on each bond with zero mean and variance $\varkappa$,
we can average Eq.~\eqref{eq:ZR} over disorder realizations to obtain
\begin{align}
\overline{Z}^R &= Z_0^R \biggl{\langle} \hT_\tau \, \mathrm{exp} \biggl{[} 
\frac{\varkappa}{2} \sum_{(ij)} \int_0^\beta \dd \tau \, \dd \tau' \, \hS_i^{z \gamma} (\tau) \hS_j^{z \gamma} (\tau) \hS_i^{z \delta} (\tau') \hS_j^{z \delta} (\tau') \biggr{]} \biggr{\rangle} \nonumber \\
&= Z_0^R \biggl{\langle} \hT_\tau \, 
\mathrm{exp} \biggl{[} 
\frac{\varkappa}{4} \int_0^\beta \dd \tau \, \dd \tau' \, \hS_i^{z \gamma} (\tau) \hS_i^{z \delta} (\tau') K_{ij} \hS_j^{z \gamma} (\tau) \hS_j^{z \delta} (\tau') \biggr{]} \biggr{\rangle},
\label{eq:Zave}
\end{align}
where in writing the second line of Eq.~\eqref{eq:Zave} we have introduced the matrix ${K}$ whose entries are $K_{ij} = 1$ if sites $i$ and $j$ are first neighbors, and $0$ otherwise. 

We now decouple the quartic product of spin operators in the exponent of Eq.~\eqref{eq:Zave} using the Hubbard-Stratonovich transformation:
\begin{equation}
\overline{Z}^R = Z_0^R \int \cD q \, \mathrm{exp} \biggl{[} -\frac{\varkappa}{4} \int_0^\beta \dd \tau \, \dd \tau' \, q_{i \tau \tau'}^{\gamma \delta} K_{i j} q_{j \tau \tau'}^{\gamma \delta} \biggr{]} \biggl{\langle} \hT_\tau \, \mathrm{exp} \biggl{[} \frac{\varkappa}{2} \int_0^\beta \dd \tau \, \dd \tau' \, q_{i \tau \tau'}^{\gamma \delta} K_{ij} \hS_j^{z \gamma} (\tau) \hS_j^{z \delta} (\tau')
 \biggr{]} \biggr{\rangle}.
 \label{eq:ZHS}
\end{equation}
To evaluate the thermal average in Eq.~\eqref{eq:ZHS}, we assume that the glass transition is of the second order and
consider the system close to the transition. Hence, the glass order parameter is small.
Employing a replica-symmetric, mean-field ansatz for the order parameter $q_{i \tau \tau'}^{\gamma \delta} \rightarrow q$, we expand the term in the angular brackets in Eq.~(\ref{eq:ZHS}) as
\begin{align}
&\quad \biggl{\langle} \hT_\tau \, \mathrm{exp} \biggl{[} \frac{\varkappa q}{2} \sum_{(\gamma \delta)} \int_0^\beta \dd \tau \, \dd \tau' \, K_{ij} \hS_j^{z \gamma} (\tau) \hS_j^{z \delta} (\tau')
 \biggr{]} \biggr{\rangle} \nonumber \\
 &= 1 + \biggl{\langle} \hT_\tau \, \frac{\varkappa q}{2} \sum_{i, (\gamma \delta)} K_{i j} \int_0^\beta \dd \tau \, \dd \tau' \, \hS_j^{z \gamma} (\tau)
\hS_j^{z \delta} (\tau') \biggr{\rangle} \nonumber \\
&+ \biggl{\langle} \hT_\tau \, \frac{\varkappa^2 q^2}{4} \sum_{i k, (\gamma_1 \delta_1), (\gamma_2 \delta_2)} K_{i j} K_{k \ell} \int_0^\beta \dd \tau \, \dd \tau' \, \dd \theta \, \dd \theta' \, \hS_j^{z \gamma_1} (\tau) \hS_j^{z \delta_1} (\tau') \hS_\ell^{z \gamma_2} (\theta) \hS_\ell^{z \delta_2} (\theta') \biggr{\rangle} + \cO (q^3),
\label{eq:q21}
\end{align}
where the now explicit replica sums run over all pairs $(\gamma \delta)$ of replicas. The $\cO (q)$ term in Eq.~\eqref{eq:q21} vanishes since clean replicas are independent of one another and the Hamiltonian $\cH_0$ is assumed to have global spin inversion symmetry. For the $\cO (q^2)$ term, the replica pair sums give two types of contributions: those in which either $\gamma_2 = \gamma_1$, $\delta_2 = \delta_1$ or $\gamma_2 = \delta_1$, $\delta_2 = \gamma_1$ and those in which there are three or four distinct replica indices. The latter type of term vanishes for the same reason as the $\cO (q)$ contribution. The former type of term, however, does not vanish, and yields
\begin{equation}
\frac{1}{4} R (R - 1) \varkappa^2 \beta^2 q^2 Z^2 N \sum_\ell \int_0^\beta \dd \tau \, \dd \tau' \, g (0, 0; \ell, \tau) g (0, 0; \ell, \tau'),
\label{eq:q22}
\end{equation}
where we define
\begin{equation}
g (j, \tau; \ell, \theta) \equiv \bigl{\langle} \hT_\tau \, \hS_j^{z \gamma} (\tau) \hS_\ell^{z \gamma} (\theta) \bigr{\rangle}.
\label{eq:g}
\end{equation}
With this ansatz, we then have, to the order $\cO (q^2)$,
\begin{equation}
\overline{Z}^R = Z_0^R + \frac{Z_0^R}{4} R (R - 1) \varkappa \beta^2 q^2 Z N \biggl{[} \varkappa Z 
\sum_\ell \int_0^\beta \dd \tau \, \dd \tau' \, g (0, 0; \ell, \tau) g (0, 0; \ell, \tau') - \frac{1}{2} \biggr{]}.
\label{eq:Zq2}
\end{equation}
We then obtain the disorder-averaged free energy functional for the glass transition to $\cO (q^2)$ as
\begin{equation}
-\beta \cF (q) = \underset{R \rightarrow 0}{\mathrm{lim}} \frac{\overline{Z}^R - 1}{R} = \frac{1}{4} \varkappa \beta^2 q^2 Z N \biggl{[} \frac{1}{2} - \varkappa Z 
\sum_\ell \int_0^\beta \dd \tau \, \dd \tau' \, g (0, 0; \ell, \tau) g (0, 0; \ell, \tau') \biggr{]} + \cO (q^3)
\label{eq:F}
\end{equation}
which yields the critical disorder strength for glass freezing:
\begin{equation}
\varkappa_c^{-1} (\beta) = 2 Z \sum_\ell \int_0^\beta \dd \tau \, \dd \tau' \, g (0, 0; \ell, \tau) g (0, 0; \ell, \tau').
\label{eq:J0critical}
\end{equation}


\subsection{Critical disorder strength in the effective Ising regime}


In what immediately follows, 
we analyse the phase boundary~\eqref{eq:J0critical} in the temperature
interval $T^* \ll T \ll \theta_W$. For simplicity, to capture the behaviour of the magnetic materials qualitatively correctly, it is sufficient to consider the case of small $\alpha\ll 1$. 

For describing the contribution of the excitations with energies $T^*\ll E\ll \theta_W$
in Eq.~\eqref{eq:J0critical}, 
we can neglect the small transverse spin-exchange-induced splitting
between states adiabatically connected to Ising ground states and consider 
the respective states to be approximately equally thermally populated.
Quantum fluctuations are thus washed out, and the system is in the effective Ising regime.
Also, the excitations of the Ising Hamiltonian with energies of order $\theta_W$ can be neglected.

We can write the $\tau$-integral over the correlation function~\eqref{eq:g} in the form
\begin{equation}
\int_0^\beta \dd \tau \, g (0, 0; \ell, \tau) = \frac{1}{Z_0 (\beta)} \sum_\bk \sum_{\bk'} \frac{\EE^{-\beta \xi_{\bk}'} - \EE^{-\beta \xi_\bk}}{\xi_\bk - \xi_{\bk'}} \langle \bk | \hS_\ell^z | \bk' \rangle \langle \bk' | \hS_0^z | \bk \rangle,
\label{eq:gint}
\end{equation}
where $\{ | \bk \rangle \}$ and $\{ \xi_\bk \}$ are respectively the eigenstates and eigenenergies of a single clean replica (we have dropped the replica indices, as they are no longer important). As discussed above, both the clean partition function $Z_0 (\beta)$ and the sum in Eq.~\eqref{eq:gint} will be dominated by the low-energy excitations 
adiabatically connected to the ground state manifold of the corresponding Ising model. In particular, the partition function can be approximated as $Z_0 (\beta) \approx \dG$, where $d_G$ is the ground state degeneracy of the Ising system, and we can make the approximation
\begin{equation}
\frac{\EE^{-\beta \xi_{\bk}'} - \EE^{-\beta \xi_\bk}}{\xi_\bk - \xi_{\bk'}} \approx \beta.
\end{equation}
Hence, Eq.~\eqref{eq:J0critical} reduces to
\begin{equation}
\varkappa_c^{-1} (\beta) \approx \frac{2 Z \beta^2}{\dG^2} \sum_\ell \biggl{(} 
\sum_{\bk_0} \sum_{\bk_0'} \langle \bk_0 | \hS_\ell^z | \bk_0' \rangle \langle \bk_0' | \hS_0^z | \bk_0 \rangle \biggr{)}^2,
\label{eq:J0Ising1}
\end{equation}
where the subscript $0$ on the state indices denotes a state which is adiabatically connected to an Ising ground state. We can decompose Eq.~\eqref{eq:J0Ising1} further as follows: writing
\begin{equation}
\sum_{\bk_0'} | \bk_0' \rangle \langle \bk_0' | = \Id - \proj_E
\end{equation}
where $\proj_E$ projects onto the excitations of $\cH_0$ which are adiabatically connected to the excited states of the Ising model, Eq.~\eqref{eq:J0Ising1} becomes
\begin{align}
\varkappa_c^{-1} (\beta) &\approx \frac{2 Z \beta^2}{d_G^2} \biggl{(} d_G^2 - 2 d_G \sum_{\bk_0} \langle \bk_0 | \hS_0^z \proj_E \hS_0^z | \bk_0 \rangle + \biggl{[} \sum_{\bk_0} \langle \bk_0 | \hS_0^z \proj_E \hS_0^z | \bk_0 \rangle \biggr{]}^2 \nonumber \\
&+ \sum_{\ell > 0} \biggl{[} \biggl{\{} \sum_{\bk_0} \langle \bk_0 | \hS_\ell^z \hS_0^z | \bk_0 \rangle \biggr{\}}^2 - 2 \sum_{\bk_0} \sum_{\bk_0'} \langle \bk_0 | \hS_\ell^z \hS_0^z | \bk_0 \rangle \langle \bk_0' | \hS_0^z \proj_E \hS_0^z | \bk_0' \rangle + \biggl{\{} \sum_{\bk_0} \langle \bk_0 | \hS_\ell^z \proj_E \hS_0^z | \bk_0 \rangle \biggr{\}}^2 \biggr{]} \biggr{)} \nonumber \\
&= 2 Z \beta^2 \biggl{(} 1 + \frac{1}{d_G^2} \sum_{\ell > 0} \biggl{[} \sum_{\bk_0} \langle \bk_0 | \hS_\ell^z \hS_0^z | \bk_0 \rangle \biggr{]}^2 \biggr{)},
\label{eq:J0Ising2}
\end{align}
where in writing the last line of Eq.~\eqref{eq:J0Ising2} we have used that since Ising ground states are eigenstates of $\hS_0^z$, and $\proj_E$ projects onto superpositions of Ising \textit{excited} states, $\proj_E \hS_0^z | \bk_0 \rangle$ vanishes. The second term in the last line of Eq.~\eqref{eq:J0Ising2} is a lattice-dependent factor of order unity. Thus, in the effective Ising regime, $\varkappa_c^{-1} \sim 2 Z / T^2$.

\subsection{Critical disorder strength at very low temperatures}
We now consider Eq.~\eqref{eq:J0critical} at low temperatures, $T \ll T^*$. In this limit, Eq.~\eqref{eq:gint} reduces to
\begin{equation}
2 \sum_{\bk} \frac{1}{\xi_{\bk}} \mathrm{Re} [\langle \bk | \hS_\ell^z | 0 \rangle \langle 0 | \hS_0^z | \bk \rangle],
\end{equation}
assuming the Hamiltonian $\cH_0$ has a non-degenerate ground state. 

Focusing on the matrix element $\langle \bk | \hS_\ell^z | 0 \rangle$, it can be shown that the many-body states of an $N$-site lattice with periodic boundary conditions take the form
\begin{equation}
\mathrm{exp} \biggl{(} \frac{i}{N} \bk \cdot \sum_{j = 1}^N \bR_j \biggr{)} | u \rangle,
\label{eq:NbodyBloch}
\end{equation}
similarly to Bloch's states in the single-particle case,
where $| u \rangle$ is a state invariant under simultaneous translation of all lattice sites by a lattice vector. Hence, defining $\hT_\ell$ as the translation operator which translates states by the lattice vector $\bR_\ell$ connecting sites $\ell$ and $0$, we have
\begin{equation}
\langle \bk | \hS_\ell^z | 0 \rangle = \langle \bk | \hT_\ell \hS_0^z \hT_\ell^\dagger | 0 \rangle = \EE^{i \bk \cdot \bR_\ell} \langle \bk | \hS_0^z | 0 \rangle.
\label{eq:matelshift}
\end{equation}
Hence, for Eq.~\eqref{eq:J0critical}, we have
\begin{equation}
\varkappa_c^{-1} = 8 Z \sum_\ell \sum_{\bk} \sum_{\bk'} \frac{\EE^{i \bR_\ell \cdot (\bk - \bk')}}{\xi_{\bk} \xi_{\bk'}} | \langle \bk | \hS_0^z | 0 \rangle |^2 |\langle \bk' | \hS_0^z | 0 \rangle |^2 = 8 Z N \sum_{\bk} \frac{1}{\xi_{\bk}^2} |\langle \bk | \hS_0^z | 0 \rangle |^4.
\label{eq:J0low1}
\end{equation}
Now, since the non-vanishing contributions to the expression~\eqref{eq:J0low1} are from the states $\{ | \bk \rangle \}$ that are superpositions of Ising ground states, whose degeneracy scales exponentially with $N$ in GF systems, the quantity $| \langle \bk | \hS_0^z | 0 \rangle |^2$ admits an expansion in powers of $N^{-1}$ with coefficients of order unity for each $\bk$. Thus, in the thermodynamic limit, the critical disorder strength is given by
\begin{equation}
\varkappa_c^{-1} \sim \frac{8 Z}{N} \sum_{\bk} \frac{1}{\xi_{\bk}^2} \rightarrow 8 Z \int \frac{\rho (\xi)}{\xi^2} \, \dd \xi
\label{eq:J0low2}
\end{equation}
where again the integral in Eq.~\eqref{eq:J0low2} is taken over the excitations which are connected to the Ising ground state manifold, i.e. the states giving rise to the low-temperature peak in the specific heat.

The critical disorder strength $\varkappa_c$ may be zero or finite depending on whether the integral in the right-hand-side of Eq.~\eqref{eq:J0low2} diverges or converges. 
The convergence of this integral is determined by the scaling of the
low-energy density of states in a particular system and by whether the excitations are gapped in the thermodynamic limit.

If the excitations remain gapped in the thermodynamic limit, the integral converges, and $\varkappa_c$ remains finite as $T \rightarrow 0$. If, however, the excitations are gapless, the integral will converge for $\rho (\xi)$ scaling faster than linearly at small $\xi$, and diverge otherwise. This means that in a gapless $3$D system with the power-law excitation dispersion $\xi (k) \propto k^p$ in the long-wave limit $k \rightarrow 0$,
\begin{equation}
\varkappa_c (T = 0) = 
    \begin{cases}
    &0, \quad p \geq 3 / 2 \\
    &A (T^*)^2 / Z, \quad p < 3 / 2
    \end{cases}
    \label{eq:J0low4}
\end{equation}
where $A$ in the second line is a lattice-dependent factor of order unity. 

Equation~\eqref{eq:J0low4} suggests that in 3D systems with linearly dispersive excitations, which are expected, e.g., on the pyrochlore lattice~\cite{SavaryBalents:review}, and in systems with a constant density of states,
the critical disorder strength $\varkappa_c (T = 0)$ at zero temperature is finite. By contrast, quadratic dispersion will lead to the vanishing of $\varkappa_c (T = 0)$.
Detailed analyses of the critical disorder strength for particular materials will be presented in future work.

\putbib
\end{bibunit}


\begin{thebibliography}{60}
\bibitem [{pyr()}]{pyrochlorenote}%
  \BibitemOpen
  \href@noop {} {}\bibinfo {note} {We note the absence of the pyrochlore lattice from the summary of the numerical results in Table~\ref{tab:C}, which we omit because the nature of its heat capacity at low temperatures is not as well-established. For instance, in Ref.~\cite{DerzhkoHutak:PHAF}, it is shown that the widely-used finite-temperature Lanczos method yields multiple peaks, while a so-called ``entropy method'' yields only a single peak.}\BibitemShut {Stop}%
\end{thebibliography}
\end{document}